# New Li-Ethylenediamine-Intercalated Superconductor Li$_x$(C$_2$H$_8$N$_2$)$_y$Fe$_{2-z}$Se$_2$ with $T_c$ = 45 K

Takehiro Hatakeda, Takashi Noji, Takayuki Kawamata, Masatsune Kato, and Yoji Koike

*Department of Applied Physics, Tohoku University, 6-6-05 Aoba, Aramaki, Aoba-ku, Sendai 980-8579, Japan*

A new iron-based superconductor Li$_x$(C$_2$H$_8$N$_2$)$_y$Fe$_{2-z}$Se$_2$ with $T_c$ = 45 K has successfully been synthesized via intercalation of dissolved lithium metal in ethylenediamine. The distance between neighboring Fe layers is 10.37 Å and much larger than those of FeSe with $T_c$ = 8 K and K$_x$Fe$_2$Se$_2$ with $T_c$ ~ 30 K. It seems that the high-$T_c$ of Li$_x$(C$_2$H$_8$N$_2$)$_y$Fe$_{2-z}$Se$_2$ is caused by the possible two-dimensional electronic structure due to the large *c*-axis length.

*E-mail: noji@teion.apph.tohoku.ac.jp

Shortly after the discovery of iron-based pnictide superconductors,[1] the iron-based chalcogenide FeSe with the PbO-type structure was reported to exhibit superconductivity with $T_c$ = 8 K by Hsu *et al.*[2] FeSe has attracted great interest, because the crystal structure is composed of a stack of edge-sharing FeSe$_4$-tetrahedra layers, which are similar to FeAs$_4$-tetrahedra layers in the iron-based pnictide superconductors, and is simpler than those of the iron-based pnictide superconductors. $T_c$'s of a FeSe family with the PbO-type structure are not so high as those of the iron-based pnictide superconductors, but the discovery of potassium-intercalated K$_x$Fe$_2$Se$_2$ with $T_c$ ~ 30 K by Guo *et al.*[3] was epoch-making. Recently, moreover, it has been reported that the intercalation into FeSe of alkali and alkaline-earth metals in liquid ammonia yields a variety of compounds with significantly enhanced $T_c$'s of 40 - 46 K.[4-6] For example, the *c*-axis length is as large as 16.518 Å in Li$_{0.9}$Fe$_2$Se$_2$(NH$_3$)$_{0.5}$ and 16.9980 Å in Li$_{1.8}$Fe$_2$Se$_2$(NH$_3$)[Li(NH$_2$)]$_{0.5}$.[5] These values are larger than 14.0367 Å in K$_x$Fe$_2$Se$_2$.[3] Very recently, Krzton-Maziopa *et al.*[7] have reported that superconductivity with $T_c$ = 45 K is induced in Li$_x$(C$_5$H$_5$N)$_y$Fe$_{2-z}$Se$_2$ via intercalation of dissolved alkaline metal in anhydrous pyridine, C$_5$H$_5$N, and that post-annealing of the intercalated material drastically expands the *c*-axis length from 16.0549 Å to 23.09648 Å and increases the superconducting shielding volume fraction.

Here, we report on the successful synthesis of a new superconductor Li$_x$(C$_2$H$_8$N$_2$)$_y$Fe$_{2-z}$Se$_2$ with $T_c$ = 45 K via intercalation of dissolved lithium metal in ethylenediamine (EDA), C$_2$H$_8$N$_2$. Post-annealing effects on the crystal structure and superconductivity are also discussed.

Polycrystalline samples of FeSe were prepared by the solid-state reaction method. Starting materials were powders of Fe and Se, which were weighted stoichiometrically, mixed thoroughly and pressed into pellets. The pellets were sealed in an evacuated quartz tube and heated at 800°C for 40 h. The obtained pellets of FeSe were pulverized into powder to use for the intercalation. Dissolved lithium metal in EDA was intercalated into the powdery FeSe as follows. An appropriate amount of the powdery FeSe was placed in a beaker filled with 0.2 M solution of pure lithium metal dissolved in EDA. The amount of FeSe was calculated in the molar ratio of Li : FeSe = 1 : 2. The reaction was carried out at 45℃ for 7 days. The product was washed with fresh EDA. All the process was performed in an argon-filled glove box. Post-annealing of as-intercalated samples was carried out at 100 – 300℃ for 60 h in an evacuated glass tube. Both FeSe and the intercalated samples were characterized by powder x-ray diffraction using Cu K$_α$ radiation of a D8 Advance Bruker AXS diffractometer. For the intercalated samples, an airtight sample holder was used. The diffraction patterns were analyzed using RIETAN – FP.[8] In order to observe the superconducting transition, the magnetic susceptibility, $\chi$, was measured using a superconducting quantum interference device (SQUID) magnetometer (Quantum Design, Model MPMS). Measurements of the electrical resistivity, $\rho$, were also carried out by the standard dc four-probe method. For the $\rho$ measurements, as-intercalated powdery samples were pressed into pellets. Then, the pellets were sintered at 200℃ for 20 h in an evacuated glass tube.

Figure 1 shows powder x-ray diffraction patterns of the as-intercalated, post-annealed (150℃, 60 h) and sintered (200℃, 20 h) samples. The broad peak around $2\theta = 20°$ is due to the airtight sample holder. Although there are unknown peaks, most of sharp Bragg peaks are due to the intercalation compound of Li$_x$(C$_2$H$_8$N$_2$)$_y$Fe$_{2-z}$Se$_2$ and the host compound of FeSe, so that they are able to be indexed based on the ThCr$_2$Si$_2$-type and PbO-type structures, respectively. Therefore, it is found that lithium and EDA are partially intercalated into FeSe, while there remains a non-intercalated region of FeSe in the samples. The lattice constants of Li$_x$(C$_2$H$_8$N$_2$)$_y$Fe$_{2-z}$Se$_2$ are calculated to be $a$ = 3.458(6) Å and $c$ = 20.74(7) Å in the as-intercalated sample. As listed in Table I, the $a$-axis length is a little smaller than that of FeSe, while the $c$-axis length is as large as that of Li$_x$(C$_5$H$_5$N)$_y$Fe$_{2-z}$Se$_2$.[7] Since the unit cell of Li$_x$(C$_2$H$_8$N$_2$)$_y$Fe$_{2-z}$Se$_2$ includes two Fe layers, the distance between neighboring Fe layers is 10.37(4) Å and much larger than 5.515(1) Å of FeSe. Our previous results have revealed that the intercalation of only lithium into Fe(Se,Te) has neither effect on the superconductivity nor crystal structure.[9] Accordingly, it is concluded that not only lithium but also EDA has been intercalated between the Se-Se layers of FeSe. As shown in Fig. 1, in fact, the simulated x-ray diffraction pattern of $RbFe_2Se_2$ with the same lattice constants, where the number of electrons of Rb in the unit cell is the same as that of $Li(C_2H_8N_2)$, is in good agreement with the diffraction pattern of $Li_x(C_2H_8N_2)_yFe_{2-z}Se_2$, suggesting that both lithium and EDA are located in the crystal site similar to that of Rb in $RbFe_2Se_2$. In addition, we have already confirmed that no intercalation of only EDA into FeSe occurs in EDA without lithium metal, whereas the intercalation of only EDA into TiNCl occurs so that superconductivity with $T_c$ = 10.5 K is induced.[10]

In Fig. 1, it is also found that the powder x-ray diffraction pattern does not change so much through the post-annealing and sintering. However, it is noticed that the (002) peak splits into two a little bit for the post-annealed and sintered samples. This suggests that another intercalated phase with a little longer $c$-axis length is induced through the post-annealing and sintering. The lattice constants are listed in Table I.

Figure 2 shows the temperature dependence of $\chi$ in a magnetic field of 10 Oe on zero-field cooling (ZFC) and on field cooling (FC) for as-intercalated, post-annealed (150℃, 60 h) and sintered (200℃, 20 h) powdery samples consisting of $Li_x(C_2H_8N_2)_yFe_{2-z}Se_2$ and FeSe. As for the as-intercalated sample, the first superconducting transition is observed at 45 K and the second transition is at 8 K. Taking into account the x-ray diffraction results, it is concluded that the first is due to bulk superconductivity of $Li_x(C_2H_8N_2)_yFe_{2-z}Se_2$, while the second is due to that of the non-intercalated region of FeSe. The superconducting volume fraction of $Li_x(C_2H_8N_2)_yFe_{2-z}Se_2$ is simply estimated as about 20 % from the ZFC measurements. The positive value of $\chi$ seems to be due to magnetic impurities taken into the sample in the intercalation process.

As for the post-annealed sample, it has been found that samples post-annealed at 100

– 200 ℃ show the first superconducting transition above 30 K, while the first superconducting transition disappears in samples post-annealed above 250℃. The $T_c$ has been found to decrease with increasing annealing-temperature. As shown in Fig. 2, $T_c$ of the sample post-annealed at 150℃ for 60 h is 41 K and is lower than that of the as-intercalated one, while the second transition at 8 K due to the non-intercalated region of FeSe is still observed. To see the temperature dependence of $\chi$ in detail, the third superconducting transition is observed at 18 K. This may be related to the additional intercalated phase with a little longer $c$-axis length as mentioned above. In any case, the post-annealing effect on the superconductivity is different from that in $Li_x(C_5H_5N)_yFe_{2-z}Se_2$.[7] This may be ascribed to the difference of the post-annealing effect on the crystal structure: that is, the lattice constants do not change so much through the post-annealing and sintering in $Li_x(C_2H_8N_2)_yFe_{2-z}Se_2$ as listed in Table I, while the $c$-axis length is markedly expanded in $Li_x(C_5H_5N)_yFe_{2-z}Se_2$.[7] This may be due to the difference of the arrangement of atoms in the molecules: that is, the arrangement of atoms in EDA is rather linear, while atoms in anhydrous pyridine form a ring similar to the benzene ring.

In order to observe the zero-resistivity, the as-intercalated powdery sample was pelletized. As shown in Fig. 3, the onset of the superconducting transition is observed at 44 K in the pellet, but zero-resistivity is not observed, which may be due to the insulating grain-boundary. Then, the pellet was sintered at 200℃ for 20 h. As shown in Fig. 3, the resistivity of the sintered pellet sample starts to decrease at 43 K with decreasing temperature and reaches zero at 18 K. The transition width is broad, which may be due to the post-annealing effect as also seen in Fig. 2.

In summary, we have succeeded in synthesizing a new intercalation compound $Li_x(C_2H_8N_2)_yFe_{2-z}Se_2$ via intercalation of dissolved lithium metal in EDA. Although the sample includes the non-intercalated region of FeSe, the $c$-axis of $Li_x(C_2H_8N_2)_yFe_{2-z}Se_2$ has drastically been expanded to 20.74(7) Å, indicating that not only lithium but also EDA is intercalated into the sample. Bulk superconductivity of $Li_x(C_2H_8N_2)_yFe_{2-z}Se_2$ has been observed below 45 K in the $\chi$ measurements of the as-intercalated sample, and moreover, zero-resistivity has been observed below 18 K for the sintered pellet sample. Compared with $T_c$ = 8 K of FeSe and $T_c$ ~ 30 K of $K_xFe_2Se_2$, $T_c$ of $Li_x(C_2H_8N_2)_yFe_{2-z}Se_2$ is clearly enhanced and is in close agreement with $T_c$'s of $Li_xFe_2Se_2(NH_3)_y$ and $Li_x(C_5H_5N)_yFe_{2-z}Se_2$ with large $c$-axis lengths. Taking into account our previous results that both the $c$-axis length and $T_c$ of Li-intercalated $Li_xFeSe$ are the same as those of FeSe,[9] it seems that the high-$T_c$ of $Li_x(C_2H_8N_2)_yFe_{2-z}Se_2$ is caused by the possible two-dimensional electronic structure due to the large $c$-axis length.

Fig. 1. (Color online) Powder x-ray diffraction patterns of the as-intercalated, post-annealed (150℃, 60 h) and sintered (200℃, 20 h) samples consisting of Li$_x$(C$_2$H$_8$N$_2$)$_y$Fe$_{2-z}$Se$_2$ and FeSe using Cu K$_\alpha$ radiation. For reference, the simulated x-ray diffraction pattern of RbFe$_2$Se$_2$ with the ThCr$_2$Si$_2$-type structure of $a$ = 3.458 Å, $c$ = 20.74 Å and the powder x-ray diffraction pattern of the host compound of FeSe are also shown. Indexes without and with asterisk are based on the ThCr$_2$Si$_2$-type and PbO-type structures, respectively. Peaks marked by ▽ and ▼ are due to Fe$_7$Se$_8$ and unknown compounds, respectively.

Fig. 2. (Color online) Temperature dependence of the magnetic susceptibility, $\chi$, in a magnetic field of 10 Oe on zero-field cooling (ZFC) and field cooling (FC) for as-intercalated, post-annealed (150℃, 60 h) and sintered (200℃, 20 h) powdery samples consisting of Li$_x$(C$_2$H$_8$N$_2$)$_y$Fe$_{2-z}$Se$_2$ and FeSe.

Fig. 3. (Color online) Temperature dependence of the electrical resistivity, $\rho$, for as-intercalated and sintered (200℃, 20 h) pellet samples consisting of Li$_x$(C$_2$H$_8$N$_2$)$_y$Fe$_{2-z}$Se$_2$ and FeSe. The inset shows the temperature dependence of $\rho$ around $T_c$.

Table I. Lattice constants $a$ and $c$ of the host compound of FeSe, FeSe and $Li_x(C_2H_8N_2)_yFe_{2-z}Se_2$ in the as-intercalated, post-annealed (150℃, 60 h) and sintered (200℃, 20 h) samples. For the post-annealed and sintered samples, two sets of lattice constants are described in correspondence to the split of the (002) diffraction peak in Fig. 1. It is noted that the distance between neighboring Fe layers of $Li_x(C_2H_8N_2)_yFe_{2-z}Se_2$ is given by the half of the $c$-axis length.

|  | $a$ (Å) | $c$ (Å) |
| --- | --- | --- |
| FeSe (host) | 3.774(1) | 5.515(1) |
| FeSe (as-intercalated sample) | 3.761(3) | 5.50(1) |
| FeSe (post-annealed sample (150 ℃, 60 h)) | 3.762(5) | 5.50(2) |
| FeSe (sintered sample (200 ℃, 20 h)) | 3.764(8) | 5.51(3) |
| $Li_x(C_2H_8N_2)_yFe_{2-z}Se_2$ (as-intercalated sample) | 3.458(6) | 20.74(7) |
| $Li_x(C_2H_8N_2)_yFe_{2-z}Se_2$ (post-annealed sample (150 ℃, 60 h)) | 3.445(4)<br>3.438(1) | 20.75(5)<br>20.92(1) |
| $Li_x(C_2H_8N_2)_yFe_{2-z}Se_2$ (sintered sample (200 ℃, 20 h)) | 3.442(4)<br>3.436(1) | 20.77(5)<br>20.92(1) |

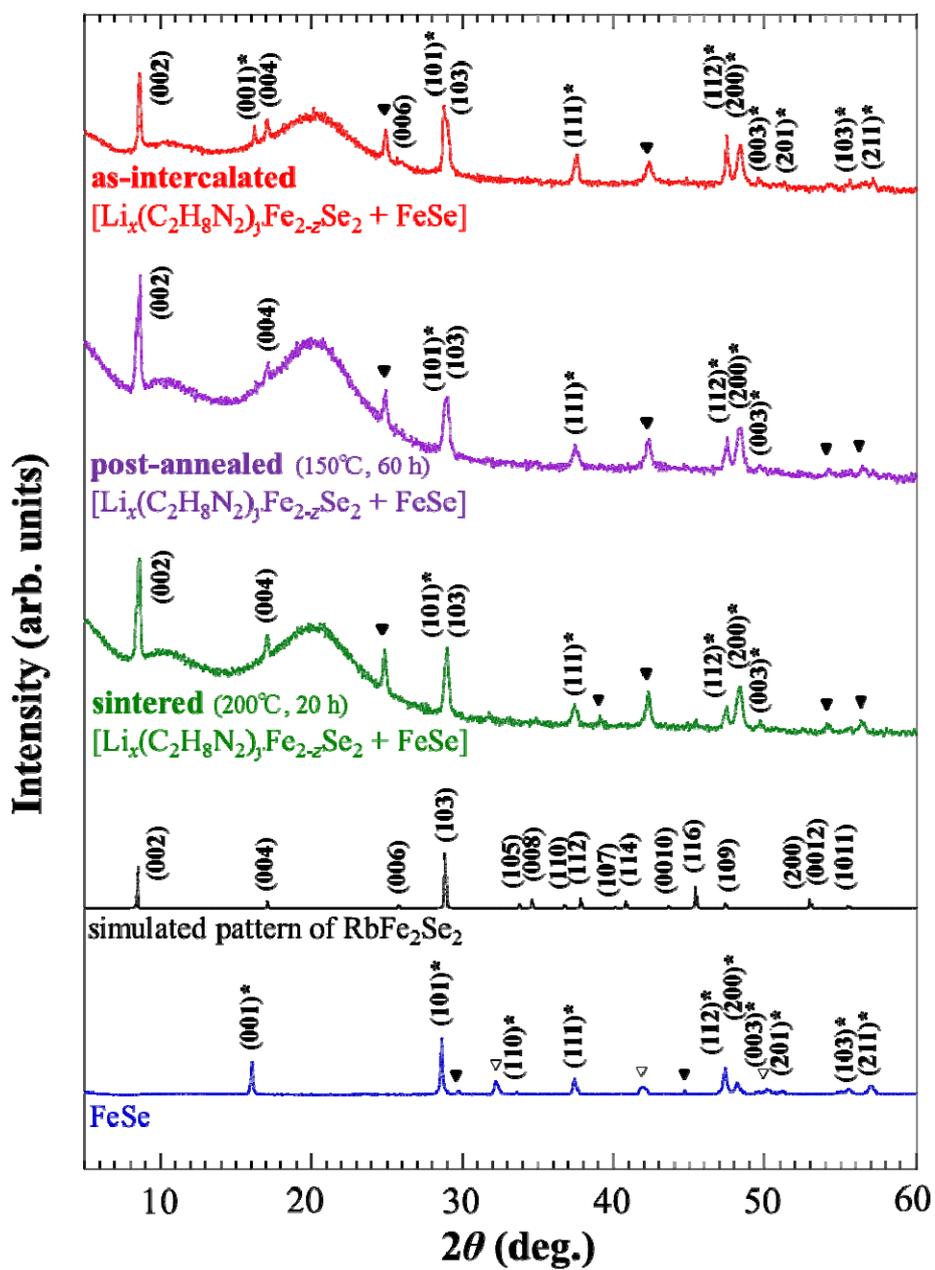

Fig. 1

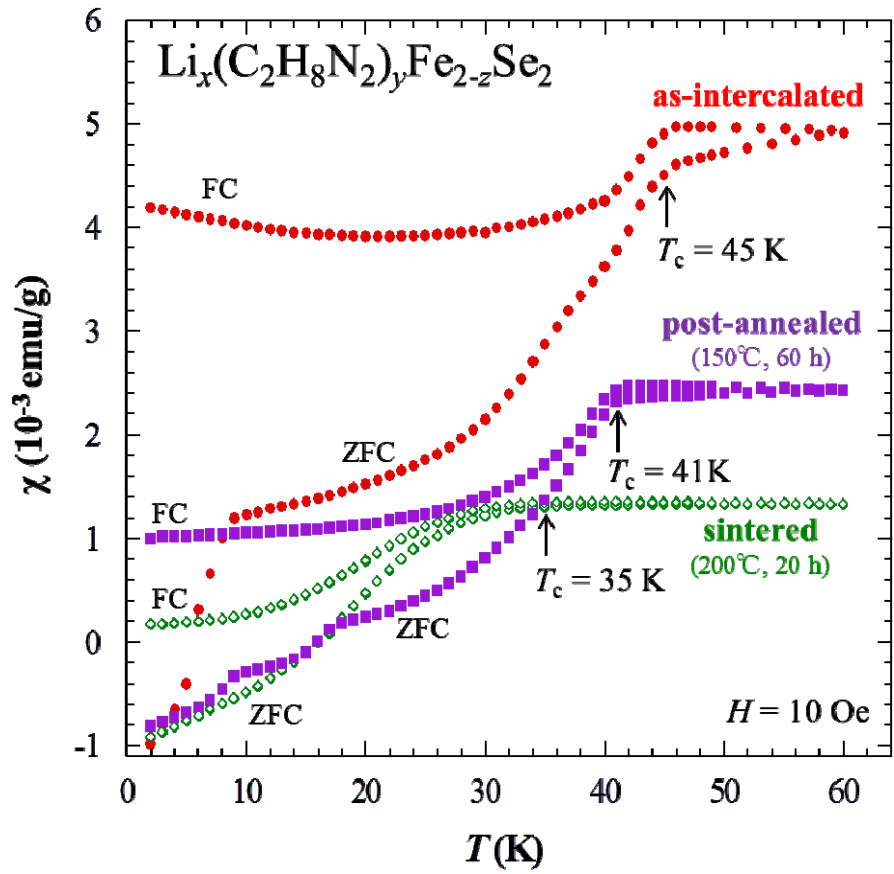

Fig. 2

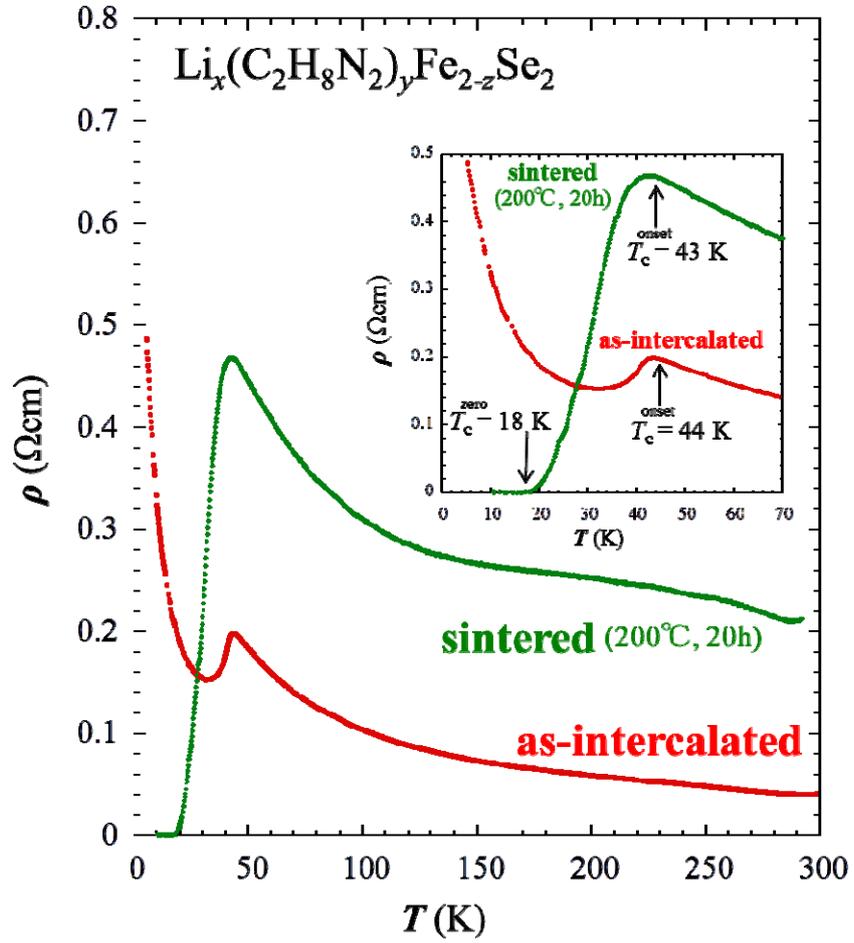

Fig. 3